\documentclass[12pt] {conm-p-l} 

\newtheorem{theorem}{Theorem}[section] 

\newtheorem{prop}{Proposition}[section]
\newtheorem{property}{Property}
\newtheorem{claim}{Claim}

\theoremstyle{definition}

\theoremstyle{remark}
\newtheorem{remark}[theorem]{Remark}

\numberwithin{equation}{section}

\newcommand{\mcl}{\mskip-10mu}   

\usepackage{epsfig}      
\usepackage{graphics}  

\copyrightinfo{2000}
    {American Mathematical Society}

\begin{document}

\title[Proof of Guesswork Necessary to Quantum Physics]{\vspace*{-60pt}A
Proof That
Measured Data and Equations of Quantum Mechanics Can Be Linked Only by
Guesswork}

\author{John M. Myers}
\address{Gordon McKay Laboratory, Harvard University, Cambridge, MA
02138, USA}
\email{myers@deas.harvard.edu}

\author{F. Hadi Madjid}
\address{82 Powers Road, Concord, MA 01742, USA}

\date{February 29, 2000}


\begin{abstract}

The design and operation of a quantum-mechanical device as a
laboratory instrument puts models written in equations of quantum
mechanics in contact with instruments.  In designing a
quantum-mechanical device of high precision, such as a quantum
computer, a scientist faces choices of models and of instruments, and
the scientist must choose which model to link to which arrangement of
instruments.  This contact is recordable in files of a Classical
Digital Process-control Computer (CPC) used both to calculate with the
equations and to manage the instruments.  By noticing that equations
and instruments make contact in a CPC, we rewrite equations of quantum
mechanics to explicitly include functions of CPC-commands to the
instruments.  This sets up a proof that a scientist's choice in
linking mathematical models to instruments is unresolvable without
guesswork to narrow the set of models from which one is to be chosen.

The proof presents the challenge of pursuing its implications.
Scientists in any investigative endeavor inherit choices from the past
and frame choices for the future, choices open to guesswork and
visible in CPC files.  To picture the framing of choices and relations
among them, we adapt colored Petri nets.  Constraining the events of
the nets to produce output colors defined by definite functions of
input colors excludes guesswork from the firing of net events, and by
contrast highlights guesses entering a net fragment as colored tokens
placed by a scientist or by instruments on input conditions.  The
availability of these net fragments makes choice and guesswork part and
parcel of physics.

Net fragments as a means of expressing guess-demanding choic\-es are
applied to portray guesswork needed in testing and calibrating a
quantum computer. The sample size required to test a quantum gate in a
quantum computer is shown to grow as the inverse square of the error
allowed in implementing the gate.

\end{abstract} 

\maketitle

\section{Introduction}

This paper stems from earlier work \cite{hj97} and a proof presented
here showing that inquiry in quantum physics continually presents a
scientist with choices of equations and of instruments, unresolvable
by calculation and measurement.  Something else is demanded of the
scientist, which may as well be called a {\em guess}.\footnote{Other
words are {\em hypothesis, Ansatz, assumption, axiom, postulate}, and
sometimes {\em principle}.}  Challenged by the proof to look at its
implications, we noticed that people in investigative endeavors
inherit and frame choices open to guesswork, some of which show up
clearly in the computers used in the endeavors.

Section 2 introduces the Classical Process-control Computer (CPC) with
its special capacity to manipulate abstractions expressed as equations
without contaminating them with its own physics.\footnote{This
capacity stems from regenerative amplifiers and clock-gated memory
registers, two inventions used to make all computer hardware
insensitive to manufacturing variations, so that, like the placing of
a chess piece not quite in the center of a square, deviations in
performance, within limits, do not matter.  Its independence from its
own physics distinguishes a classical computer from a quantum
computer.}  A scientist can use a CPC not only to calculate with
equations, but also to mediate the command of laboratory instruments
via digital/analog (D/A) converters and to record experimental results
returned from the instruments via analog/digital (A/D) converters.  By
noticing that both equations and instruments make contact in a CPC, we
rewrite equations of quantum mechanics to explicitly include functions
of CPC-commands to the instruments.  This sets up the proof that the
scientist's choice in linking equations to instruments is unresolvable
without guesswork to narrow the set of models.  A lattice of sets of
models is defined, two widely used guesses that narrow the set of
models are noted, and the concept of statistical distance between
probability distributions is applied to quantum-mechanical models.

Section 3 provides language for displaying and analyzing
guess-demanding choices visible in CPC files.  To this end Turing
machines are introduced and adapted to formalize the definition of a
CPC.  This allows fragments of colored Petri-nets, opened to exogenous
influences, to portray the programming and running of programs in a
network of CPC's operated by collaborating scientists.  Many of these
programs incorporate guesses.  This general picture of process-control
computation shows programs and other guesses as colors on tokens that
a scientist enters on a Petri net that acts as a game board.
Mechanisms for one scientist to judge programs (and hence guesses)
made by another are sketched, leading to the first of many needs for
concurrently operating CPC's.

Section 4 describes some examples in which guessing, quantum mechanics
and CPC structures must interact in the building of a quantum computer
as a laboratory instrument specified by equations of quantum
mechanics.  We show the need for guesses to link equations and
instruments brings with it a need to test the quantum computing
instruments and to calibrate them, guided by test results and guesses.
Quantum mechanics imposes a peculiar structure on this testing,
related to the statistical distance between models.  For the measure
of precision conventionally used in quantum computing, the sample size
needed for testing a quantum gate is shown to increase as the inverse
square of the tolerated imprecision.  While many questions are left
open to future work, the example demonstrates a frame for analysis and
experiment broader than any quantum model alone, a frame that includes
the testing of the mathematical models by results of the use of
instruments, and so distinguishes what the model says from what the
instruments do, allowing provision for guesswork as an ingredient in
advancing both models and instruments.

\section{Quantum-mechanical models and their links to instruments}

Proving the necessity of guesses demands language to describe the
linking of numbers in mathematical models to numbers pertaining to
laboratory instruments, starting with mathematical language to
describe a scientist's choosing one arrangement of laboratory
instruments rather than another.  We shall describe a situation in
which a scientist chooses instruments by using a CPC keyboard to type
strings of characters, much as G\"odel, in mathematical logic,
described equations as strings of characters.  The scientist at the
CPC keyboard writes and executes programs to command the operation of
laboratory instruments and record their results.  These programs make
use of quantum-mechanical equations, which the scientist also writes
into the CPC.

Quantum mechanics as a mathematical language expresses different
measuring instruments by different operators, and thus has built into
it a recognition that phenomena to be described cannot be independent
of the instruments used to study them \cite{dirac}.  Still, this
dependence is emphasized more some times than others.  Some modeling
merely assumes that instruments can be found, without saying how, to
implement various combinations of state vectors and operators.  Such
models appear in theories of quantum computing to relate the
multiplication of unitary operators to the solving of problems of
interest.  To see the need for another kind of model, suppose a
scientist has computer-controlled instruments (such as lasers) with
the potential to implement a quantum computer, and faces the question
of what commands the CPC should transmit to the instruments and when
it should transmit them in order implement one or another quantum
gate.  Determining the commands and their timing to implement a
quantum gate expressed as a unitary matrix $U_j$ takes a model that
expresses the gates as unitary transformations in terms of commands
that a process-control computer can transmit to the instruments.
Curiously, models of this kind have not been much stressed in physics,
and it is a merit of efforts to build quantum computers to make
the importance of such models apparent.

\subsection{Models and instruments make contact in a CPC}

Part of a scientist's control of instruments can work through the use
of a process-control computer that transmits commands to the
(computer-controlled) instruments and records results produced by
them.  We confine our analysis to this part, excluding from
consideration here (but by no means denigrating) hand work beyond the
reach of a process-control computer.  We shall portray cases in which
a scientist chooses arrangements of instruments, chooses models, and
puts the two in contact, linking models to instruments, during a CPC
session starting after the instruments have been set up and put under
control of a CPC and ending before the scientist has to tinker with
the instruments in ways unreachable by the CPC.  Within the CPC,
laboratory instruments and mathematical models make contact when:
\begin{enumerate} \item a model resident in a CPC file is used to
derive commands for the CPC to transmit to the instruments; \item
instrumental results collected by a CPC are used to narrow down a set
of models.  (We shall later see feedback as an example of this.)
\end{enumerate} Such contact does not spring from nothing, but is
brought about by design and depends on choices made by a scientist,
including choices of what set of models to start with, what model to
choose for use by a CPC in generating commands, and what experiments
to run.  To picture the design and operation of contact between models
and instruments, imagine eavesdropping on CPC's used in various
investigations.  Commands sent to the instruments by the CPC and the
results received from them, both numerical, are amenable to analysis,
as is the scientist's writing of equations, programs, calls for
program execution, etc.; we also eavesdrop on displays produced by the
CPC for the scientist.

Although the CPC puts instruments in contact with equations involving
quantum superposition, the CPC itself is a classical machine, free of
quantum superposition, for it needs no quantum behavior within itself,
neither to manipulate equations of quantum mechanics nor to manage
laboratory instruments.  For example, the writing of an expression
$|0\rangle + |1\rangle$ for a superposition of quantum states makes
use of written characters that themselves exhibit no superposition.
And any command to instruments is likewise a character string, including a
command to rotate a polarizing filter by 45 degrees to implement the
superposed state $|0\rangle + |1\rangle$.  Similarly, results of the use
of instruments interpreted as demonstrating superposition arrive as
bit strings, themselves devoid of superposition.

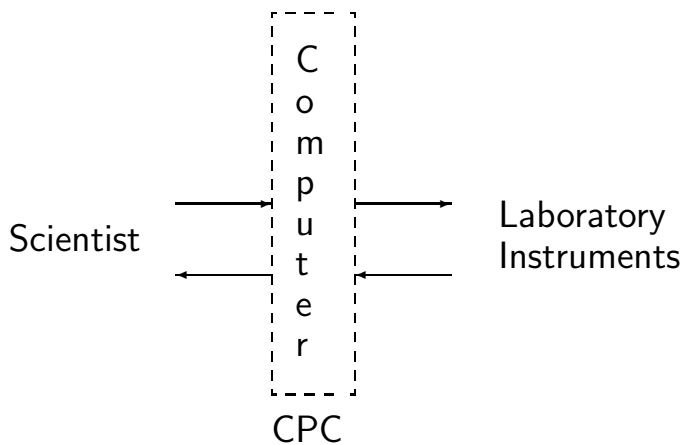
\begin{figure}[tb]
\hskip-60pt\setlength{\unitlength}{3947sp}%
\begingroup\makeatletter\ifx\SetFigFont\undefined%
\gdef\SetFigFont#1#2#3#4#5{%
  \reset@font\fontsize{#1}{#2pt}%
  \fontfamily{#3}\fontseries{#4}\fontshape{#5}%
  \selectfont}%
\fi\endgroup%
\begin{picture}(3075,2712)(1726,-2536)
\thinlines
\put(2776,-1036){\vector( 1, 0){600}}
\put(3376,-1486){\vector(-1, 0){600}}
\put(3901,-1036){\vector( 1, 0){600}}
\put(4501,-1486){\vector(-1, 0){600}}
\put(3376,-2236){\dashbox{60}(525,2400){}}
\put(4801,-1186){\makebox(0,0)[lb]{\smash{\SetFigFont{14}{16.8}{\sfdefault}{\mddefault}{\updefault}Laboratory}}}
\put(4801,-1441){\makebox(0,0)[lb]{\smash{\SetFigFont{14}{16.8}{\sfdefault}{\mddefault}{\updefault}Instruments}}}
\put(3526,-211){\makebox(0,0)[lb]{\smash{\SetFigFont{14}{16.8}{\sfdefault}{\mddefault}{\updefault}C}}}
\put(3526,-466){\makebox(0,0)[lb]{\smash{\SetFigFont{14}{16.8}{\sfdefault}{\mddefault}{\updefault}o}}}
\put(3526,-721){\makebox(0,0)[lb]{\smash{\SetFigFont{14}{16.8}{\sfdefault}{\mddefault}{\updefault}m}}}
\put(3526,-976){\makebox(0,0)[lb]{\smash{\SetFigFont{14}{16.8}{\sfdefault}{\mddefault}{\updefault}p}}}
\put(3526,-1231){\makebox(0,0)[lb]{\smash{\SetFigFont{14}{16.8}{\sfdefault}{\mddefault}{\updefault}u}}}
\put(3526,-1486){\makebox(0,0)[lb]{\smash{\SetFigFont{14}{16.8}{\sfdefault}{\mddefault}{\updefault}t}}}
\put(3526,-1741){\makebox(0,0)[lb]{\smash{\SetFigFont{14}{16.8}{\sfdefault}{\mddefault}{\updefault}e}}}
\put(3526,-1996){\makebox(0,0)[lb]{\smash{\SetFigFont{14}{16.8}{\sfdefault}{\mddefault}{\updefault}r}}}
\put(1726,-1336){\makebox(0,0)[lb]{\smash{\SetFigFont{14}{16.8}{\sfdefault}{\mddefault}{\updefault}Scientist}}}
\put(3376,-2536){\makebox(0,0)[lb]{\smash{\SetFigFont{14}{16.8}{\sfdefault}{\mddefault}{\updefault}CPC}}}
\end{picture}\caption{Computer mediating contact
between scientist and instruments.}  \label{fig:sides}\end{figure}

The CPC is situated between a scientist to its left and laboratory
instruments to its right, as shown in figure \ref{fig:sides}.  Working
at the CPC, a scientist is limited in action at any moment to the
resolution of the choice presented by the CPC at that moment, a choice
defined by the files stored in its memory and the state of its
processor, and exemplified by a menu displayed by the CPC. Our analysis
of the CPC cannot reach beyond its buffers: neither to its left into
the scientist, nor to its right where, invisible to eavesdropping,
reside digital-to-analog (D/A) and analog-to-digital (A/D) converters
and beyond them the laboratory instruments.

\subsection{Quantum-mechanical models that recognize com-\break mands
sent to instruments}

For equations of quantum mechanics to model effects of a scientist's
choices in arranging instruments, these choices must show up in the
equations.  To see how this can work, recall that quantum mechanics
parses the functioning of instruments into state preparation,
transformation, and measurement, three coordinated activities that
generate {\em outcomes}, supposed visible in experimental results by
some means unspecified.  The three activities are described,
respectively, by a state (as a unit vector representing a ray in a
Hilbert space), a unitary operator, and a hermitian operator.  The
only way to make the scientist's choices in arranging instruments show
up in quantum-mechanical equations is to make the state vector
$|v\rangle$, the transformation operator $U$, or the measurement
operator $M$, or some combination of them, depend on how these choices
are resolved.

A simple and yet, so far as we know, original way to analyze a
scientist's choice of arrangements of instruments is to suppose that
during a CPC-mediated session the instruments are controlled by
CPC-transmitted commands from a set $B$ of possible commands, where $B
\subset \mathcal{B}$ and $\mathcal{B}$ is the set of all finite binary
strings.  We formulate a core set of quantum mechanical models that
express the probability of an outcome of instruments in response to a
command $b \in B$ sent to the instruments by the CPC, as follows.  Let
$\mathcal{V}_B$, $\mathcal{U}_B$, and $\mathcal{M}_B$ be the sets of
all functions $|v\rangle$, $U$, and $M$, respectively, with
\begin{eqnarray*}
|v\rangle: B &\mcl\rightarrow\mcl& \mathcal{H},\\[-3pt]
U: B &\mcl\rightarrow\mcl& \{\mbox{unitary operators on } \mathcal{H} \},\
\mbox{and}\\[-3pt]
M: B &\mcl\rightarrow\mcl& \{\mbox{hermitian operators on } \mathcal{H}
\}.\end{eqnarray*}
The core models exhibit discrete spectra for all $M \in \mathcal{M}$:
\begin{property} \begin{equation} (\forall b \in B) M(b) =
\sum_j m_j(b) M_j(b),
\label{eq:basic} \end{equation}
where $m_j: B \rightarrow \mathcal{R}$ (with $\mathcal{R}$ denoting
the real numbers) is the $j$-th eigenvalue of $M$, and $M_j$ is the
projection onto the $j$-th eigenspace (so $M_j M_k = \delta_{j,k}M_j$).
\end{property}  Let $\Pr(j|b)$ denote the probability of obtaining
the $j$-th outcome, given transmission by the CPC of a command $b$.
Although not commonly seen in texts, this probability of an outcome
given a command is the hinge pin for focusing on quantum mechanical
modeling of uses of instruments.  Quantum mechanics constrains the
models to satisfy:
\begin{property}
\begin{equation}  \Pr(j|b) = \langle
v(b)|U^\dagger(b)M_j(b)U(b)|v(b)\rangle , \end{equation} where the
$\dagger$ denotes the hermitian adjoint. \end{property} (Within this
modeling scheme, the Schr\"odinger equation relates a model at a later
time to a model at an earlier time by a certain transformation
operator $U$, dependent on the situation.)

Any choice from the sets $\mathcal{B}$, $\mathcal{V}_B$,
$\mathcal{U}_B$, and $\mathcal{M}_B$ produces some quantum-mechan\-ical
model $(|v\rangle, U, M)_B$.  Two models $(|v\rangle, U, M)_B$ and
$(|v'\rangle, U', M')_B$ generate the same probabilities $\Pr(j|b)$ if
they are unitarily equivalent, meaning there exists a $Q: B
\rightarrow \{\mbox{unitary operators on }$ $\mathcal{H} \}$ such that
$(\forall b \in B) |v'(b)\rangle = Q(b)|v(b)\rangle$, $U'(b) = Q(b)
U(b) Q^\dagger(b)$ and $M'(b) = Q(b) M(b) Q^\dagger(b)$.  For this
reason, any model $(|v\rangle, U, M)_B$ can be reduced to
$(|v'\rangle, {\bf 1}, M)_B$, where $|v'\rangle = U|v\rangle$
and $M' = M$ or, alternatively to $(|v\rangle, {\bf 1}, M')_B$ where
$M' = U^\dagger M U$.

More models are available in more general formulations.  When we show
that guesswork is necessary even to resolve choices among models of
the core set, it follow that guesswork is necessary also to resolve
the choices of among a larger set of models involving
positive-operator-valued measures, superoperators, etc.

\subsection{From results to quantum-mechanical outcomes}

Before stating and proving the proposition that calculations and
measurements cannot by themselves link models to outcomes obtained
from instruments, we call to the reader's attention that outcomes
themselves, in the sense of quantum mechanics, are produced by
instruments only with the help of interpretive guesswork.

\begin{claim} To speak of actual instruments in the language of
quantum mechanics one needs to associate results of the use of the
instruments, recorded in a CPC, with {\em outcomes} in the sense of
quantum mechanics or with averages of outcomes. \end{claim}

Experimental results of the use of instruments become
quantum-mechanical outcomes only by a scientist's act of interpreting
the results as outcomes.  The interpretation involves judgment and
guesswork, not only to sidestep imperfections in the instruments, but
as a matter of principle, even for the limiting case of instruments
supposed free of imperfections.  For example, light detectors used in
experiments described by models of quantum optics generate
experimental results; typically, each of $L$ detectors reports
to the CPC at each of a succession of $K$ time intervals a detection
result, consisting of 0 (for no detection) or 1 (for detection), so a
record contains $LK$ bits.  Depending on judgments made about
correlations from time interval to time interval and detector to
detector, these $LK$ bits may constitute $LK$ quantum outcomes, or one
quantum outcome, or some number in between.  The number of outcomes in
$LK$ bits is determined neither by the experimental results (which in
this case are just these bits) nor by general principles of quantum
mechanics; yet the parsing of results into outcomes must occur, at
least provisionally, before any comparison between equations and
measured outcomes can be made.  Henceforth, when we speak of outcomes,
we presuppose that this piece of guesswork has been accomplished and a
decision made to define the parsing of results into outcomes.

\subsection{Calculation and measurement by themselves cannot link
quantum models to recorded outcomes}

Could it be that the general properties 1 and 2 suffice to determine a
model (up to unitary equivalence) if only one collects enough measured
results interpreted as outcomes?  The answer is: ``no; unless some
special properties restrict the model more tightly than the form
established by properties 1 and 2 alone, one can always find many
unitarily inequivalent models $(|v\rangle, U, M)_B$, all
of which produce probabilities that match perfectly the relative
frequencies of outcomes.''

To prove this we define some things to pose the issue more sharply.
Let $B$ denote the set of commands used to generate some set of
outcomes interpreted from measured results.\footnote{Practically
speaking, $B$ must be a finite set, but the proof holds also for $B$
denumerably infinite.}  For any $b \in B$, let $N(b)$ be the number
of times that an outcome has been entered in the record for a run of
the experiment for command $b$, and let $J(b)$ be the number of
distinct outcomes for command $b$.  For $j = 1, \ldots , J(b)$,
let $\lambda_j(b)$ be the $j$-th distinct outcome obtained for command
$b$, and let $n(j,b)$ be the number of times this $j$-th distinct
outcome $\lambda_j$ is recorded in response to command $b$.  For all
$j > J(b)$ let $\mu_j(b)$ be arbitrary real numbers, and for all $j
\geq 1$ let $\phi(j,b)$ be arbitrary real numbers.

\begin{prop}
Given any set of recorded outcomes associated with any set $B$ of
commands, the set of models satisfying properties 1 and 2 contains
many unitarily inequivalent models $(|v\rangle, U, M)_B$, each of which
has a perfect fit with the set of outcomes, in the sense that
\begin{equation}(\forall b)(\forall 1 \leq j \leq J(b)) \Pr(j|b) =
n(j,b)/N(b). \end{equation} \end{prop}
\begin{proof}
It is instructive to start with the special case in which for
some $b \in B$, there exist two or more distinct values of $j$ for
which $n(j) > 0$.  For this case let the set $\{ |j\rangle \}$ be an
orthonormal basis of a separable Hilbert space.  Define a subset $S$
of models satisfying properties 1 and 2 as all models of the form
$(|v\rangle, U, M)_B$, where \begin{eqnarray} |v(b)\rangle &\mcl
\stackrel{\rm def}{=}\mcl&
\sum_{j=1}^{J(b)}[n(j,b)/N(b)]^{1/2}\exp(i\phi(j,b)|j\rangle), \\ U(b)
&\mcl\stackrel{\rm def}{=}\mcl& {\bf 1},\\ M(b) &\mcl\stackrel{\rm def}{=}
\mcl& \sum_{j=1}^{J(b)}\lambda_j(b)|j\rangle \langle j| +
\sum_{j=J(b)+1}^\infty\mu_j(b)|j\rangle \langle j|, \end{eqnarray}
with $\mu_j$ and $\phi$ arbitrary real-valued functions.  By invoking
property 2, one checks that any such model has the claimed perfect
fit; yet the set contains many unitarily inequivalent models, which
predict conflicting statistics for some possible quantum
measurement.\footnote{This happens {\em e.g.}\ for primed and unprimed
models if for any $b$, $n(1,b) \neq 0 \neq n(2,b)$ and $\phi(1,b)-\phi(2,b)
\neq \phi'(1,b)-\phi'(2,b)$.} This proves the special case.

For the general case, modify the definitions above to
\begin{eqnarray} |v(b)\rangle &\mcl
\stackrel{\rm def}{=}\mcl&
\sum_{j=1}^{J(b)}[n(j,b)/N(b)]^{1/2}|w_j\rangle, \\
U(b) &\mcl\stackrel{\rm def}{=}\mcl& {\bf 1},\\ M(b)
&\mcl\stackrel{\rm def}{=}\mcl& \sum_{j=1}^{J(b)}\lambda_j(b)P_j +
\sum_{j=J(b)+1}^\infty\mu_j(b) P_j, \end{eqnarray} where $P_j P_k =
\delta_{j,k}P_j$, for all $j$ the projection $P_j$ has dimension
greater than 1, and $|w_j \rangle$ ranges over all unit vectors of the
eigenspace defined by $P_j |w_j\rangle = |w_j \rangle$.  In
particular, for any $j$, dim($P_j$) can be as large as one pleases.
Then even if there is only one outcome that is ever recorded, there
are still many unitarily inequivalent models that perfectly fit the
data.\end{proof}

Proposition 2.1 implies that the density matrix, often supposed to be
determined from measured data \cite{belt}, is undetermined without
assuming special properties shortly to be discussed; this follows by
expressing the density matrix as $|v\rangle \langle v|$ and noticing
that the phases of the off-diagonal elements are undetermined.  We
leave to the future the demonstration of additional ambiguity in the
link between any set of recorded outcomes and models expressed in the
mathematical language of quantum mechanics.

\subsection{Statistically significant differences between models}

In practice, a scientist has little interest in a model chosen so that
its probabilities exactly fit measured relative frequencies.  Rather,
the scientist wants a simpler model with some appealing structure that
comes reasonably close to fitting.  Quantum mechanics encourages this
predilection, because on account of statistical variation in the
sample mean, functions that perfectly fit outcomes on hand at one time
are not apt to fit perfectly outcomes acquired subsequently.  We
show here that accepting statistics no way takes away from the proof
that measurements and equations by themselves cannot link models to
instruments.

One needs a criterion for the statistical significance of a difference
between two quantum-mechanical models (or between a model and measured
relative frequencies).  Here we limit our attention to models $\alpha$
and $\beta$ which have a set $B$ of commands in common and for which
the spectra of $M_\alpha$ and $M_\beta$ are the same.  For a single
command $b$, the question is whether the difference between the
probability distributions $\Pr_\alpha(\cdot|b)$ and
$\Pr_\beta(\cdot|b)$ is bigger than typical fluctuations expected in
$N(B)$ trials.  An answer is that two distributions are
indistinguishable statistically in $N(b)$ trials unless
\begin{equation} N(b)^{1/2}d(\mbox{$\Pr_\alpha$}(\cdot|b),
\mbox{$\Pr_\beta$}(\cdot|b)) > 1, \label{eq:dist} \end{equation} where
$d$ is the statistical distance defined by Wooters in Eq.\ (10) of
\cite{wooters}.  Furthermore, Wooters's Eq.\ (12) shows for two models
$\alpha$ and $\beta$ that differ only in the function $|v\rangle$,
\begin{equation} d(\mbox{$\Pr_\alpha$}(\cdot|b),
\mbox{$\Pr_\beta$}(\cdot|b)) \leq \cos^{-1}|\langle v_\alpha(b)|
v_\beta(b) \rangle |. \label{eq:dangle}
\end{equation}

To judge the significance of the difference between two models
with respect to a set B of commands common to them, a scientist who
chooses some weighting of different commands can define a weighted
average of $d(\Pr_\alpha(\cdot|b), \Pr_\beta(\cdot|b))$ over all $b
\in B$.  The same holds if model $\beta$ is replaced by relative
frequencies of outcomes interpreted from measured results.

It is noteworthy that the set of models statistically
indistinguishable from a given model can be much larger than
would be the case if the ``$\leq$'' of (\ref{eq:dangle}) were an
equality, as follows.

\begin{prop} For any set of outcomes, two models $\alpha$ and $\beta$ 
of the form $(|v\rangle,{\bf 1},M)_B$ can perfectly fit the relative
frequencies of the outcomes (Proposition 2.1) and yet be mutually
orthogonal in the sense that $\langle v_\alpha| v_\beta \rangle = 0$
\end{prop}
\begin{proof}
For any set of measured outcomes, there exists a perfectly fitting
model $\alpha$ of the form in the proof of Proposition 2.1 for the
general case, for which $(\forall j,b) dim(|w_j(b) \rangle ) > 1)$,
and a corresponding perfectly fitting model $\beta$ such that
$(\forall j,b) \langle w_{\alpha,j}(b)| w_{\beta,j}(b)\rangle = 0$. For
these two models, $\langle v_(b)\alpha(b)| v_\beta(b)\rangle =
\sum_j [n(j,b)/N(b)] \langle w_{\alpha,j}(b)| w_{\beta,j}(b)\rangle =
0$ . \end{proof}

Wooters extended the definition of statistical difference to unit
vectors.  While for any two unit vectors, there exist measurement
operators that maximize the statistical distance between them, for
any such operator there exist other vectors, mutually orthogonal, that
have zero statistical distance relative to this operator.  For this
reason, among others, statistics still leaves the scientist needing
something beyond calculation and measurement to determine a model, for
the set of models closer than $\epsilon$ in weighted statistical
distance to certain measured results certainly includes all the models
that exactly fit the data and, without special restrictions dependent
on guesses, this set includes models that are mutually orthogonal.
Models close to given measured data are not necessarily close to
each other in the predictions they make.

\subsection{Lattices of models}

Properties 1 and 2 set up a big set of models $(|v\rangle,U,M)_B$, $B
\subseteq \mathcal {B}$, $|v\rangle \in \mathcal{V}, U \in
\mathcal{U}, M \in \mathcal{M}$.  Subsets of models of this set
are a lattice under set intersection and union.  Each command set $B$
establishes a smaller lattice of sets of models, and these lattices
will play a part in the testing and calibrating of quantum computers,
discussed in section 4, where a scientist encountering problems with a
model chooses a set of possible alternatives, and then tries to narrow
it.  Often this narrowing is seen as choosing values of parameters
within a form of model in order to obtain a best fit, say with a
criterion of minimizing statistical distance between frequencies of
outcomes interpreted from measured results and probabilities
calculated from the model.  One is free to think of the estimating of
parameters in the language of a lattice of models as the using
of measured results to select a model from a set of models.

From Proposition 2.1 that showed that the whole set of models defined
by properties 1 and 2 is too big to permit measured results to select
a model, we have:
\begin{prop} For measured data to uniquely decide to within unitary
equivalence which quantum-me\-chan\-ical model of a set of models best
fits experimental results interpreted as outcomes by a criterion of
least statistical distance (or any other plausible criterion), the set
of models must first be sufficiently narrowed, and this narrowing is
underivable from the results and the basic properties 1 and 2 of
quantum mechanics. \end{prop}

Something beyond measured results and calculations from equations is
required to narrow a set of models so that measured results can select
a model that is ``best'' by some criterion.  Such an act of choosing
undefined by calculation and results of observation is what we have
called a {\em guess}.

\subsection{Hidden guesswork in conventional quantum mechanical models}
The proof casts in a clear light maneuvers conventionally made to
narrow down the set of models.  Sometimes a community of physicists is
in mutual agreement about guesses deemed appropriate, and this
agreement obscures from notice the fact that a guess is invoked.  As
an example of a widely invoked guess, most modeling in quantum physics
supposes that the scientist can vary $b$ so as to vary $U(b)$ while
holding $v(b)$ and $M(b)$ constant.  Indeed, most models used in
quantum physics are restricted to the subset of models having the
special
\begin{property} The command $b$ is the
concatenation of separate commands for the three types of operations, so
that
\begin{equation}
b = b_v\parallel b_U\parallel b_M \label{eq:bcat}, \end{equation}
where here the $\parallel$ denotes concatenation of commands.
\end{property}
According to these models, one can vary any one of the three while
holding the other two fixed.  This specializes (\ref{eq:basic})
to the more restrictive form:
\begin{equation}
\Pr(j|b) = \langle v(b_v)|U^\dagger(b_U)M_j(b_M)U(b_U)|v(b_v)\rangle .
\label{eq:basic2} \end{equation}

An additional constraining guess characterizes models widely used in
the analysis of quantum computers, a guess prompted by the desire to
generate a unitary transformation as a product of other unitary
transformations that serve as ``elementary quantum gates.''  For
example, one may want to generate the unitary transformation
$U(b_{U,1})$ $U(b_{U,2})$.  To generate it one causes the CPC to transmit
some $b_U$.  For quantum computing to have an advantage over classical
computing, the determination of this $b_U$ in terms of $b_{U,1}$ and
$b_{U,2}$ must be of polynomial complexity \cite{complex}.  It is
usually assumed that $b_U$ is the simplest possible function of
$b_{U,1}$ and $b_{U,2}$, as follows.

Let $B_U \subset B$ be a set of instrument-controlling commands,
thought of as strings that can be concatenated.  Suppose the function
$U$ has the form $U(b_1\parallel b_2) = U(b_2)U(b_1)$ for all $b_1
\parallel b_2 \in {B_U}$ (note reversal of order).  Then we say the
function $U$ respects concatenation.

\begin{property} Quantum computation employs a subset of models in
which $U$ respects concatenation.  \end{property}

\begin{remark}
We present properties 1 through 4 not as properties of laboratory
instruments, but as properties that a scientist can choose to demand
of models.  Whether the instruments act that way is another question.
There are reasons, relaxation and other forms of decoherence among
them, to expect limits to the precision with which instruments can
behave in accord with properties 3 and 4.  All four properties are
used often enough to be conventions, in the sense that a convention is
a guess endorsed by a community.
\end{remark}

\section{Petri nets to show choices open to guesswork}

In orchestrating contact between mathematical models and laboratory
instruments, scientists set up chains of cause and effect, expressed
in computer programs with their ``if-then'' structure, not as static
propositions but as designs for action.  Such designs are implemented
in experiments; an example is a feedback loop that adjusts the
orientation of a filter according to a rule that tells what adjustment
to make in immediate response to a result recorded by a light
detector.  On a more relaxed time scale, physicists make other
connections by analyzing outcomes of one generation of experiment,
using the equations of a model, to set up design instruments for a
next generation.  As remarked above, contact between equations and
instruments depends on choices made by scientists, including choices
of what set of models to start with, what model to choose for use by a
CPC in generating commands, and what experiments to run.  If these
choices could be resolved by some combination of calculation and
measurement, one could argue that they are irrelevant to physics.  But
the propositions of the preceding section show this is not the case,
so the design and operation of contact between equations and
instruments, with its ineradicable dependence on guesswork, cries out
for attention as part and parcel of physics.

Although widespread in practice, the design of contact between
equations and instruments is in its infancy as a topic for theoretical
attention.  A beginning can be seen in Benioff's analysis of sequences
of measurements (described quantum mechanically) in which subsequent
measurements are functions of outcomes of preceding
measurements \cite{benioff}.  Called decision procedures, these involve
classical feedback control equations to control instruments described
quantum mechanically, in some cases with proved advantages \cite{hj93}.
These efforts dealt with measurements occurring at a single location.
Designs that put equations and instruments in contact over a network
of cooperating investigators are wide open for future attention.

Logic in experiments, in feedback loops at many time scales, is logic
in action.  This is the logic of models that relate instrument
commands to quantum vectors and operators.  Here we adapt Petri nets
to provide mathematical language by which to express and analyze
designs for contact between equations and instruments, designs that
include sequencing of effects, decision rules, and interactions among
sequences of effects that scientists implement in their instruments.
The nets will highlight choices resolvable only by resort to
guesswork; they serve as a language with which one can express
formally how guessing works in physics, case by case, within
CPC-mediated investigations.

\subsection{Requirements CPC's}

In order to adapt Petri nets to showing guess-demanding choices
visible in CPC's, we start by clarifying how a CPC differs from a
Turing machine, on the way to adapting the Turing machine to process
control and to use in a network of collaborating scientists.  This
lays the groundwork for introducing Petri nets.

\subsubsection{Timing in the execution of commands} 
The first thing that makes process-control computing special is
timing.  In the context of quantum-mechanical models, each unitary
transformation maps states possible in one situation to states
possible in another situation; for quantum computing this means
mapping states possible at an earlier time to states possible at a
later time.  Thus a unitary transformation is implemented not all at
once, but over a time duration.  In practice, that duration depends on
how the instruments implement the transformation.  A written command
$b_U$ acts as a musical score.  Like sight reading at a piano,
executing a program containing the command $b_U$ requires converting
the character string $b_U$---the score---into precisely timed
actions---the music.  The piano keys, in this analogy, include the output
buffers that control the amplitude, phase, frequency, and polarization
of lasers of an ion-trap quantum computer or of radio-frequency
transmitters for a nuclear-magnetic-resonance (NMR) quantum computer.

For this reason executing a command $b_U$ requires parsing it into
pieces (signals) and implementing each signal at a time, the
specification of which is contained in the string $b_U$. Either the
CPC that executes a program in which $b_U$ is written parses the
command into signals and transmits each signal at its appointed time,
or the instruments receiving the command $b_U$, unparsed, contain
programmable counters operating in conjunction with a clock that do
this timed parsing.  Such programmable counters themselves constitute
a special-purpose CPC.  So either the scientist's CPC must execute
commands by issuing an appropriately timed sequence of signals, or
some other CPC attached to the instruments must do this.  Either way,
the capacity to execute programmed motion in step with a clock is a
requirement for a CPC, distinct from and in addition to requirements
to act as a Turing machine.

\subsubsection{Firewalls in a network of computers}

Just as axioms set up branches of mathematics, guesses set up rules for the
conduct of experiments and the interpretation of their results, rules
often embedded in CPC's.  Collaborating scientists accept guesses
from each other, at least provisionally, use these in experimenting
and modeling; they evaluate some of them, sometimes refining or
replacing them.  This poses a problem for CPC-mediated inquiry, where
guesses engender computer programs, for a scientist's guess can
reprogram a CPC, often for better but sometimes, by malice or
accident, for worse.  Scientists in a collaboration need to test each
other's programs and to limit the influence of any program, making the
scope of influence of a CPC program a matter for negotiation among the
collaborators.

An easy but narrow case is that of a computer running G\"odel's test for
validity of a claimed derivation \cite{godel}.  To think about such
testing, one models the computer by a Turing machine designed to start
from a tape on which the claimed derivation is written and to halt
leaving a ``yes'' or ``no'' on the tape, according to whether the
claim is or is not valid.  Such a Turing machine can be emulated by a
universal Turing machine executing a testing program to check a
passive (non-executed) file containing the claimed derivation.

Not just derivations, but also programs need to be tested with respect
to what they do when they are executed.  But what is to keep an
executing program under test from infecting the program that tests it?
Hardware walls of some kind are needed.  By limiting our analysis to
exclude remote login and insisting on computers that distinguish
physically one interface from another, we can see a basic structure
for testing programs and for limiting the reach of guesses of any one
scientist in a network of CPC's, based on operating two or more CPC's
concurrently with controlled interfaces between them, so the testing
program and the program under test execute on separate CPC's, with an
interface controlled by the testing CPC.  By virtue of concurrent
operation of CPC's with controlled interfaces, guesses made by
collaborators can set up programs that frame choices open to guessing
by any one scientist, and that test the performance of the scientist's
programs within that frame of choice, allowing freedom to a scientist
to program one part of the investigation while insulating other parts.
Hardware walls that limit the reach of one person's guesses at any
moment are one many motivations for stressing a network of
concurrently operation CPC's.

\subsection{Turing machines and Petri nets}

Here we provide language for displaying and analyzing guess-demanding
choices visible in files of CPC's used by collaborating scientists who
on occasion reprogram those choices.  As a model of a CPC, we assume
that each CPC of a network is a Turing Machine adapted for
Process-control (TMP), to be defined.  Making sense of networks of
TMP's handling equations and controlling instruments calls for a
descriptive capacity that allows for various viewpoints at various
levels of detail.  We introduce a specialized use of fragments of
colored Petri-nets, opened to exogenous influences, to portray the
programming and running of programs in a network of TMP's operated by
collaborating scientists.\footnote{Our use of Petri nets is
impressionistic and a more technical presentation will doubtless be
rewarded by exposing issues here overlooked.}

Different viewpoints and levels of detail are accommodated by morphisms
in the category of nets.  Isomorphisms between Petri nets trade net
detail for color detail \cite{jensen}.  These will be combined with
coarsening maps that suppress detail, for example by mapping colored
tokens to black tokens.  We will show how the programming of a
universal TMP (UTMP) portrayable as a single Petri net can produce any
number of patterns of use of instruments and equations, portrayable by
a host of different Petri nets.  This general picture of
process-control computation will show programs and other guesses as
colors on tokens that a scientist enters on a game board defined by a
fragment of a Petri net, and equations of quantum mechanics written as
guesses by a scientist will be seen as colors on tokens that take part
in directing and interpreting the use of laboratory instruments.

\subsubsection{Writing {\em vs.}\ executing a program}
Computers rest on the writing of motionless characters on a page to
describe something moving, a puzzle solved in music by writing notes
on staves, to be read in step with a swinging pendulum that chops time
into moments, so that written notes that portray a still picture for
each moment direct the motion of the playing of a musical instruments
\cite{crosby}.  The logical machinery of a computer moves in response
to triggering signals, ``tick'' and ``tock'', synchronized to distinct
phases of the swinging of a pendulum.  Computer designers employ truth
tables, each of which specifies the response of a clocked circuit at a
tock to a stimulus present as an input at a preceding tick.  A row of a
truth-table can be drawn as a transition in a Finite State Machine
(FSM).  By coupling an FSM to a memory of unlimited capacity, one
arrives at the theoretical concept of a Turing machine, various
special cases of which perform various special tasks
\cite{turing,feynman,boolos}.  And here is the crux of programming:
because a state machine is describable by still writing---a table---a
Turing machine can be designed to be universal.  By coding into its
memory the table that describes any given special Turing machine, one
causes the universal Turing machine to emulate the given special
Turing machine.  So, apart from speed and memory requirements, the
single universal Turing machine can be put to doing any of the things
that any of the special Turing machines can do, making it potentially
convenient, once adapted to process control, to designing and
implementing contact between equations and instruments.  (But demands
for quick response require in some cases devices streamlined to a
special task better modeled by a special Turing machine than by a
universal one.)  The next tasks are to adapt the Turing machines,
special and universal, to process control, and after that to express
them formally by use of colored Petri net fragments.

\subsubsection{Turing machine for process control (TMP)}
\begin{figure}[tb]
\epsfysize=4.75in\begin{center}
\epsfbox{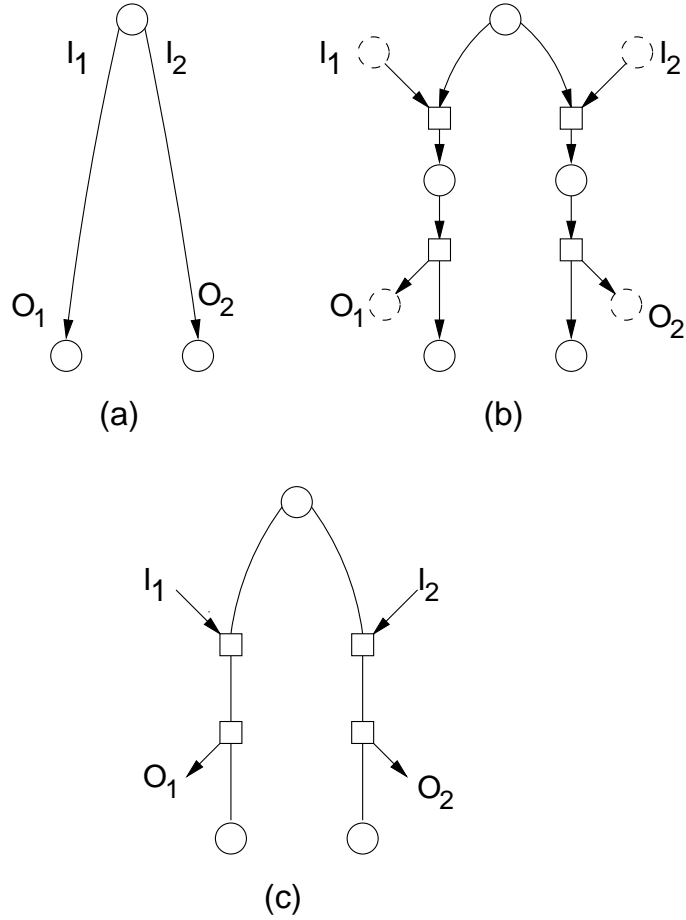}
\end{center}
\vskip-.5\baselineskip
\caption{Fragment of FSM.}
\label{fig:fsm1}\end{figure}

To adapt a Turing machine as a model of a process-control computer, we
leave the coupling of the FSM to the memory unchanged but add input
and output buffers to the FSM.  As for the FSM, at whatever level of
detail of description one chooses, the control structure of a program
(with its ``if-then'' statements) can be viewed as an FSM consisting
of (classical) states drawn as circles, connected by directed arcs,
with each arc labeled by an input $I$ that selects it and by an output
$O$ \cite{feynman}; a fragment of such a picture is shown in figure
\ref{fig:fsm1}(a).  An FSM serves as a game board on which a single
token can be placed to mark the ``current state.''  Heading toward the
hooking together of FSM's to make a Petri net, we suppose that each
arc in the FSM is punctuated by a tick event and a tock event, drawn
as small boxes, enlarging the FSM into a special case of a
condition-event Petri net fragment, as shown in figure
\ref{fig:fsm1}(b). Once colors are introduced, states shown as dashed
circles pointing into an event of the FSM from outside will become the
means to express the entrance of guesses. These states are assumed to
receive tokens put into them by scientists and instruments undescribed
by events of the net.  Similarly, dashed states pointed to by arcs
from an event are assumed to have tokens taken from them by agents
undescribed by events of the net.  Figure \ref{fig:fsm1}(c)
streamlines the picture to the form we shall use, in which more or
less vertical arcs are understood to point downward, the dashed states
are left undrawn, as are all states with one input and one output
event.  To emphasize the input and output arcs with their extra tokens,
we often call this an FSM fragment to distinguish it from the FSM form
of figure \ref{fig:fsm1}(a).

To define a Turing machine for Process-control (TMP), we adapt the
FSM of a Turing machine to have for each of its states a cartesian
product of states of a set of clocked internal registers and, in
addition, input buffers and output buffers, which allow
input/output transactions with a scientist, with laboratory
instruments, and with other TMP's.

\subsubsection{Colored tokens}

\begin{figure}[t]
\begin{center}
\epsfbox{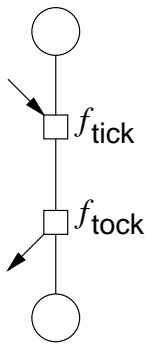}
\end{center}
\caption{FSM with detail pushed to coloring.}
\label{fig:fsm2}\end{figure}

By replacing the black tokens of an FSM fragment by a colored tokens
and adjoining to each event a function that defines colors on output
tokens in terms of colors on input tokens, any FSM fragment can be
mapped one-to-one to the drastic form of figure \ref{fig:fsm2}, in
which color changes substitute for most of the moves of black tokens
on a bigger net.  A ``fork in the road'' for black tokens, turns into
a choice between red and green, so to speak, so the descriptive burden
is taken up by the functions $f_{\mbox{tick}}$ and $f_{\mbox{tock}}$;
$f_{\mbox{tick}}$ defines the color of a token placed on an internal
state depending on a list of colors, one for each input, while
$f_{\mbox{tock}}$ defines a list of output colors depending on the
color of the token on the internal state.  The vertical arc is to be
read as directed downward, and the big circles at the top and bottom
of a path signify that the path is wrapped around a cylinder, so the
top is a continuation of the bottom, {\em i.e.}\ a loop.  An FSM
fragment in which the token carries a color will be called a colored
FSM.

\subsubsection{Other mappings}
Less drastic mappings are also possible.  Any two states of a single
FSM can be merged without breaking any arcs by augmenting
the color rules in the events that feed them and the events fed by
them.  If a set of states connected to one another by events is mapped
into a single state, the single state then connects to an event that
loops back to it; this results in a place-transition Petri net, but
not a condition-event net.  We restrict the mappings dealt with here
to ones that avoid pasting tick and tock events together, thereby
avoiding self loops.  Two events of an FSM that link the same pair of
states can be merged by distinguishing external inputs and outputs by
color instead of by place.  

\begin{figure}[t]
\begin{center}
\epsfxsize=4.3in\epsfbox{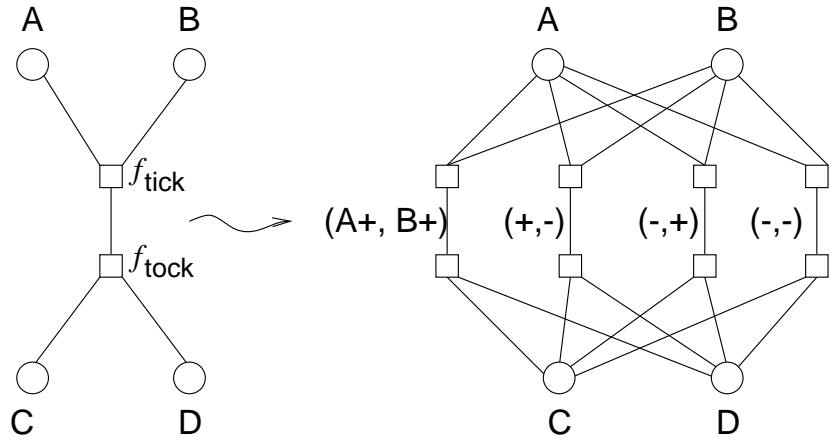}
\end{center}
\caption{From color detail to net detail.}
\label{fig:net}\end{figure}

The mappings discussed so far are net isomorphisms: they map markings
of one net bijectively to markings of the other and preserve the
one-step reachability of one marking from another (by the firing of an
event).  Inverses of these bijections take more richly to less
richly colored nets.  Going in this direction depends on each state of
a colored FSM having a set of possible colors associated with
it \cite{jensen}; then any colored transition corresponds one-to-one to a
set of transitions obtained by partitioning sets of colors of input
states, as illustrated in figure \ref{fig:net} for a two-in, two-out
transition with color sets $A$, $B$, $C$, and $D$, each partitioned
into ``+'' and ``$-$'' subsets.  For this to make sense, it must be
that an event which has tokens in all its inputs cannot fire unless
the colors of the tokens comprise an element of the domain of its color
function; we assume this firing rule.

One gets a coarser description by use of a surjective map that is not
an isomorphism by dropping the color distinction and dropping the
color functions from the transitions; this coarsening, however,
preserves a one-to-one correspondence between the number of firings in
one net and the number in another.  All these maps are continuous in
the net topology \cite{genrich}, and, as emphasized by Petri
\cite{petri1}, nets form a category in which the morphisms are
continuous maps, an idea that extends to nets with colored tokens
\cite{jensen}.

\subsubsection{Disciplined coarsening of time}

Some other kinds of continuous coarsening maps bundle up multiple
event firings into a single firing; as when one describes {\em e.g.}\
``running a program'' as a single event.  This brings us to the first
of several areas open to future work, for, more than other computing,
process control benefits from well defined timing, and in particular
from machine and software design that allows systematic, well controlled
mappings that take a certain number of firings in an FSM to a single
firing, so that one can think at a coarser level while still
maintaining discipline in timing.  

A striking example of the need to design programs that run in the same
time for all inputs from some set $I$ occurs in quantum computing. For
example, suppose that ${\bf U}$ is the universal unitary operator
defined by Deutsch to operate on basis states of the form $|s;{\bf
n};{\bf m}\rangle$ where $s$ is the location of the scanned square,
${\bf n}$ is the state of the FSM-processor (${\bf n}=0$ is the
starting state and ${\bf n}=1$ the halt state) and ${\bf m}$ is the
tape \cite{deutsch}. For this to work in a computation that takes
advantage of quantum superposition, one needs $\exists r [(\forall x
\in I) {\bf U}^r |0;0;x,0 \rangle = |0;1;x,f(x) \rangle]$; however,
this is by no means implied for a program $\pi_f$ for which (as is
usual in borrowed classical programs) one can assure only $(\forall x
\in I) (\exists r(x))[{\bf U}^{r(x)}|0;0;x, 0 \rangle = |0;1;x, f(x)
\rangle ]$ \cite{myers}.  An interesting topic for future study is the
complexity of converting various classes of programs with variable
running time to programs running in a time independent of the input
for some set of inputs.

\subsubsection{Cartoon of UTMP}
\begin{figure}[tb]
\noindent\hskip-60pt\resizebox{4.125in}{!}{\setlength{\unitlength}{3947sp}%
\begingroup\makeatletter\ifx\SetFigFont\undefined%
\gdef\SetFigFont#1#2#3#4#5{%
  \reset@font\fontsize{#1}{#2pt}%
  \fontfamily{#3}\fontseries{#4}\fontshape{#5}%
  \selectfont}%
\fi\endgroup%
\begin{picture}(5337,3393)(901,-3211)
\thinlines
\put(2701,-211){\circle{300}}
\put(1801,-2536){\circle{300}}
\put(1801,-211){\circle{300}}
\put(4201,-2536){\circle{300}}
\put(4201,-211){\circle{300}}
\put(5926,-2536){\circle{300}}
\put(5926,-211){\circle{300}}
\put(2701,-2536){\circle{300}}
\put(2701,-361){\line( 0,-1){525}}
\put(1876,-1786){\line( 1, 0){750}}
\put(1501,-661){\vector( 1,-1){225}}
\put(1726,-1861){\vector(-1,-1){225}}
\put(1876,-961){\line( 1, 0){750}}
\put(3901,-661){\vector( 1,-1){225}}
\put(4126,-1861){\vector(-1,-1){225}}
\put(5626,-661){\vector( 1,-1){225}}
\put(6226,-661){\vector(-1,-1){225}}
\put(5851,-1861){\vector(-1,-1){225}}
\put(6001,-1861){\vector( 1,-1){225}}
\put(2701,-1861){\line( 0,-1){525}}
\put(2701,-1036){\line( 0,-1){675}}
\put(1801,-361){\line( 0,-1){525}}
\put(1801,-1036){\line( 0,-1){675}}
\put(1801,-1861){\line( 0,-1){525}}
\put(4201,-361){\line( 0,-1){525}}
\put(4201,-1036){\line( 0,-1){675}}
\put(4201,-1861){\line( 0,-1){525}}
\put(5926,-361){\line( 0,-1){525}}
\put(5926,-1036){\line( 0,-1){675}}
\put(5926,-1861){\line( 0,-1){525}}
\put(1351, 14){\makebox(0,0)[lb]{\smash{\SetFigFont{14}{16.8}{\sfdefault}{\mddefault}{\updefault}FSM}}}
\put(2401, 14){\makebox(0,0)[lb]{\smash{\SetFigFont{14}{16.8}{\sfdefault}{\mddefault}{\updefault}Memory}}}
\put(901,-1411){\makebox(0,0)[lb]{\smash{\SetFigFont{14}{16.8}{\sfdefault}{\mddefault}{\updefault}Scientist}}}
\put(3901, 14){\makebox(0,0)[lb]{\smash{\SetFigFont{14}{16.8}{\sfdefault}{\mddefault}{\updefault}UTM}}}
\put(3301,-1411){\makebox(0,0)[lb]{\smash{\SetFigFont{14}{16.8}{\sfdefault}{\mddefault}{\updefault}Scientist}}}
\put(2101,-3211){\makebox(0,0)[lb]{\smash{\SetFigFont{14}{16.8}{\sfdefault}{\mddefault}{\updefault}(a)}}}
\put(4051,-3211){\makebox(0,0)[lb]{\smash{\SetFigFont{14}{16.8}{\sfdefault}{\mddefault}{\updefault}(b)}}}
\put(5776,-3211){\makebox(0,0)[lb]{\smash{\SetFigFont{14}{16.8}{\sfdefault}{\mddefault}{\updefault}(c)}}}
\put(5551, 14){\makebox(0,0)[lb]{\smash{\SetFigFont{14}{16.8}{\sfdefault}{\mddefault}{\updefault}UTMP}}}
\put(5026,-1411){\makebox(0,0)[lb]{\smash{\SetFigFont{14}{16.8}{\sfdefault}{\mddefault}{\updefault}Scientist}}}
\put(6076,-1411){\makebox(0,0)[lb]{\smash{\SetFigFont{14}{16.8}{\sfdefault}{\mddefault}{\updefault}Instruments}}}
\put(1201,-1861){\makebox(0,0)[lb]{\smash{\SetFigFont{14}{16.8}{\sfdefault}{\mddefault}{\updefault}$T_2$}}}
\put(1201,-736){\makebox(0,0)[lb]{\smash{\SetFigFont{14}{16.8}{\sfdefault}{\mddefault}{\updefault}$T_1$}}}
\put(1726,-1861){\framebox(150,150){}}
\put(2626,-1861){\framebox(150,150){}}
\put(1726,-1036){\framebox(150,150){}}
\put(2626,-1036){\framebox(150,150){}}
\put(4126,-1861){\framebox(150,150){}}
\put(4126,-1036){\framebox(150,150){}}
\put(5851,-1861){\framebox(150,150){}}
\put(5851,-1036){\framebox(150,150){}}
\end{picture}} 
\caption{From FSM to UTMP.}
\label{fig:utm}\end{figure}
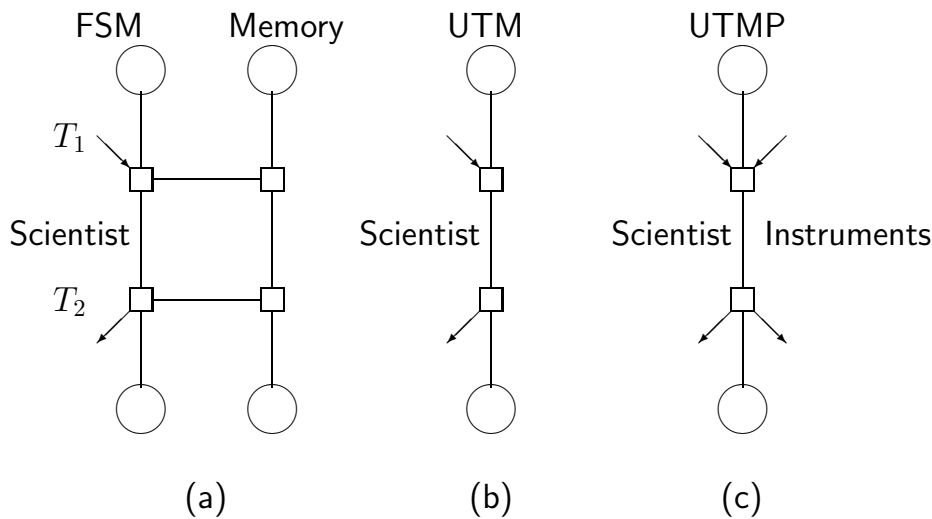

Ignoring the laboratory instruments for the moment, by connecting
input- and output-signals from a suitable FSM to a scientist and
coupling the FSM to an unlimited memory, one gets a Universal Turing
Machine (UTM) that provides for continual communication with a
scientist, as shown in figure \ref{fig:utm}(a), in which boxes
connected by a horizontal line are read as a single event.  We cartoon
the UTM in the condensed form of figure \ref{fig:utm}(b).  By adding
input- and output-signals from the FSM to laboratory instruments and
to other FSM's, one gets a Universal Turing Machine adapted for
Process control (UTMP), as shown in figure \ref{fig:utm}(c); again
almost all of the burden of description is in the color functions,
here called $T_1$ and $T_2$ (for Turing) that define a finite state
machine that operates a UTMP.  We assume that at some level of
description, the ticks and tocks of the UTMP slice time into moments
not only for the UTMP but also for the scientist at a keyboard and the
instruments on the laboratory bench; we assume that input tokens from
the scientist and from the instruments arrive at the UTMP synchronized
with the UTMP pendulum.  If the scientist enters nothing at a given
clock tick, then the token taken by the UTMP from the input buffer for
the scientist carries the color ``empty,'' and if the instruments
enter nothing, the token from the input buffer for the instruments
carries the color ``empty''; similarly the UTMP marks output tokens
with the color ``empty'' if it writes nothing else on them.

\subsubsection{A scientist controls a UTMP}
To see the structure imposed on physics by the UTMP, one must think as
if the UTMP were delivered to a scientist in a bare condition: no
installed software,\footnote{The scientist can borrow software and
install it, but is responsible for it.} the FSM in a starting state,
and the memory all blank.  We assume that the function $T_1$ operating
on empty input tokens, the starting state of the FSM, and a blank
memory produces empty output tokens and makes no change in the FSM
state or the memory or the memory location scanned.  Finally, we
invoke the universality of a UTM to assume that the functions $T_1$
and $T_2$ are fixed (by a manufacturer, so to speak) independent of
whatever laboratory instruments need to be considered and independent
of all action by the scientist.  These assumptions imply

\begin{prop}Whatever a UTMP does besides staying in its starting state
and taking in and putting out empty tokens is in response to input
tokens. \end{prop}

We invoke this proposition to view the scientist as precluded from
defending questionable management of equations or instruments by
saying ``the computer did it.''  If a CPC does something, it executes
a program; we view the scientist as responsible for any program
entered (as a colored token) into the UTMP and for running the program
on any particular occasion.\footnote{This rules out taking for granted
the operating system, instrument-managing programs, a simulator, and
whatever other programs come pre-installed in a commercially available
CPC.}

\subsubsection{Reprogramming always an option}

\begin{figure}[t]
\begin{center}
\epsfbox{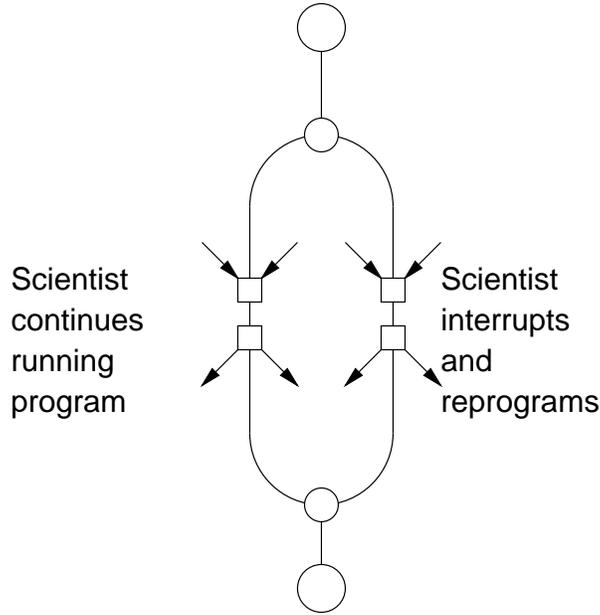}
\end{center}
\caption{Alternative modes controlled by scientist.}
\label{fig:utmp2}\end{figure}

We assume the UTMP is isomorphic to the net shown in figure
\ref{fig:utmp2}, so that the scientist has a recurring choice of
letting the UTMP run as programmed or of interrupting it to reprogram
it.  By programming a UTMP, a scientist can simulate an arbitrary
special Turing machine.  At will, the scientist can interrupt a
program in execution to change to a program that simulates a different
special Turing machine, corresponding to a different FSM and a
different net.  One can glimpse this in figure \ref{fig:net}, where it
is apparent that if the colors are limited to the sets $A^+$ and
$B^+$, then six of the eight events are precluded from firing, and the
net is in effect reduced to the fragment defined by the selected
colors.  In this way the part of the net that actually fires,
corresponding to the event ``Use existing program'' of figure
\ref{fig:utmp2}, is variable in how it acts and in the net by which
one portrays it in more detail, according to the scientist's actions
in providing and running programs.

\subsubsection{Plug and play}
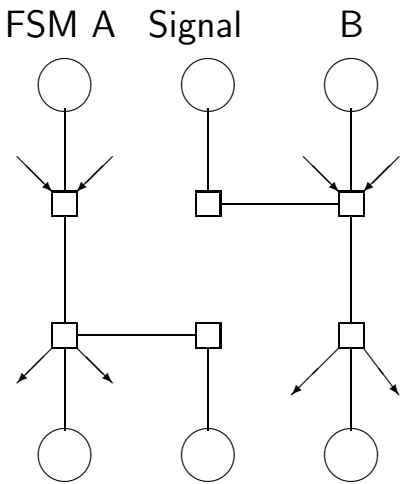
\begin{figure}[t]
\begin{center} \setlength{\unitlength}{3947sp}%
\begingroup\makeatletter\ifx\SetFigFont\undefined%
\gdef\SetFigFont#1#2#3#4#5{%
  \reset@font\fontsize{#1}{#2pt}%
  \fontfamily{#3}\fontseries{#4}\fontshape{#5}%
  \selectfont}%
\fi\endgroup%
\begin{picture}(2487,2950)(1426,-2693)
\thinlines
\put(2701,-211){\circle{300}}
\put(1801,-2536){\circle{300}}
\put(2701,-2536){\circle{300}}
\put(3601,-2536){\circle{300}}
\put(3601,-211){\circle{300}}
\put(1801,-211){\circle{300}}
\put(2701,-361){\line( 0,-1){525}}
\put(2701,-1861){\line( 0,-1){525}}
\put(2776,-961){\line( 1, 0){750}}
\put(1876,-1786){\line( 1, 0){750}}
\put(1501,-661){\vector( 1,-1){225}}
\put(2101,-661){\vector(-1,-1){225}}
\put(1726,-1861){\vector(-1,-1){225}}
\put(1876,-1861){\vector( 1,-1){225}}
\put(3301,-661){\vector( 1,-1){225}}
\put(3901,-661){\vector(-1,-1){225}}
\put(3526,-1861){\vector(-1,-1){300}}
\put(3676,-1861){\vector( 3,-4){225}}
\put(1801,-361){\line( 0,-1){525}}
\put(1801,-1036){\line( 0,-1){675}}
\put(1801,-1861){\line( 0,-1){525}}
\put(3601,-361){\line( 0,-1){525}}
\put(3601,-1036){\line( 0,-1){675}}
\put(3601,-1861){\line( 0,-1){525}}
\put(1426, 89){\makebox(0,0)[lb]{\smash{\SetFigFont{14}{16.8}{\sfdefault}{\mddefault}{\updefault}FSM A}}}
\put(2326, 89){\makebox(0,0)[lb]{\smash{\SetFigFont{14}{16.8}{\sfdefault}{\mddefault}{\updefault}Signal}}}
\put(3526, 89){\makebox(0,0)[lb]{\smash{\SetFigFont{14}{16.8}{\sfdefault}{\mddefault}{\updefault}B}}}
\put(1726,-1861){\framebox(150,150){}}
\put(2626,-1861){\framebox(150,150){}}
\put(3526,-1861){\framebox(150,150){}}
\put(1726,-1036){\framebox(150,150){}}
\put(2626,-1036){\framebox(150,150){}}
\put(3526,-1036){\framebox(150,150){}}
\end{picture} \end{center}
\caption{Signal from A to B.}
\label{fig:utm3}\end{figure}

To see how UTMP's can be connected (as well as the detail of how the
FSM of a TM or TMP is connected to the memory), we introduce a signal
that is phased just opposite to an FSM: the signal takes an input at a
tock event and issues an output at a tick event.  Then FSM A can send
a signal (which can convey a message as a token color) to B (which can
be either another FSM or a memory), as shown in figure \ref{fig:utm3},
provided the signal path is short enough compared to the clock rate of
the machines.  This use of a signal synchronizes A \hbox{with B}.  For
two-way communication, one adds a signal going the other way.  If
communication over a distance long compared to the clock period is
called for, then a chain of communication over intermediating UTMP's,
is necessary, with the result that more firings of an event of A are
required before a consequence of one firing can propagate to B and
return as a property of a color on a token at a later firing of the
A-event.  The use of colored tokens sets up an area for future
investigation of replacing the awkward definition of {\em synchronic
distance} \cite{synch} with a measure of synchronization that counts
firings in circuits of color effects, without having to add artificial
elements to a net.

\subsection{Net fragments formalized}

Portraying logic operating in CPC's calls for fragments of Petri nets,
not complete nets, to allow for guesses as token colors definable
neither by results of experiments nor by calculations.  From among the
standard definitions of a Petri net, the one we use is $(S, E, F)$,
where $S$ is a set of states, $E$ is a set of events, and $F \subseteq
S \times E \cup E \times S$ is the flow relation.  In order to make
room for guesses from a scientist and results of instruments
inexpressible in the logic defined by a net but essential to setting
it up, the nets used are all \textit{net fragments}, which we define
as follows.  A net fragment is a structure $(S, S_I, S_O, E, F)$ where
$S$ is a set of states of CPC's, and $S_I$ is a set of states of input
signals ({\em e.g.}\ from A/D converters to a CPC input buffer),
disjoint from $S$, allowing for input to the CPC from a scientist and
laboratory instruments.  $S_O$ is a set of states of output signals
disjoint from both $S$ and $S_I$, allowing for output from the CPC;
the flow relation is expanded so $F \subseteq [(S \cup S_I) \times E]
\cup [E \times (S \cup S_O)]$.  States of $S_I$ are assumed to have
tokens placed in them by some means beyond the net, and states of
$S_O$ are assumed to have tokens removed from them by means beyond the
net.  Our pictures show stubs of arcs from states of $S_I$ to events
and from events to states of $S_O$ while omitting the circles for
these states.  Associated with a net fragment is a ``reduced net''
obtained by omitting the states of $S_I$ and $S_O$ (and dropping the
arc stubs); using this reduced net, one can explore issues of liveness
and safety \cite{berthelot}.  The events of $E$ express computer logic
and nothing else.  As an example of a guess used in designing contact
between equations and instruments, a mathematical model entered by a
scientist as a colored token in an $S_I$ state can assert whatever
rules the scientist chooses to relate tokens received from instruments
in $S_I$ states to commands sent to them as colored tokens in $S_O$
states.  In this way the net fragment expresses the difference between
such a model, with its guesswork, as a color on a token and how the
instruments actually behave by producing colored tokens on their own.

\section{Net-based portrait of guesswork needed to test 
and calibrate a quantum computer}

In section 2 choices of equations to link to instruments were shown
inescapably open to guesswork, bidding to make guesswork part and
parcel of physics.  The availability of net fragments described in
section 3 brings within physics the study of contacts between
equations and instruments by making available to analysis relations of
sequence, concurrency and choice expressed in these contacts and in
the guess-dependent actions that set the contacts up.  Here we turn
from nets themselves to attention to an example problem in which a net
illustrates an important structure needed to link equations to
instruments.  Besides the net explicitly shown in figure 8, the
availability of nets provides a framework in which to view the main
topic of this section, the problem of resolving a choice of commands
by which a CPC manages a quantum computer.  That framework can be used
in the future to ask other questions, to do with: how do the necessities
of quantum-mechanical models, classical process control, and guesswork
interact; how are FSM's as program structures affected by use of
models that are quantum mechanical; how does the need for CPC's to
mediate between quantum-mechanical equations and instruments change
our understanding of quantum mechanics?

Turning to the case at hand, some telling illustrations of guesswork
needed to link models to instruments arise in quantum computing.  To
build a quantum computer, say to solve problems of factorizing
\cite{shor} and searching \cite{grover}, a scientist must choose
quantum-mechanical equations and laboratory instruments to work in
harmony.  Quantum computational models call for quantum gates that are
unitary transformations, each a tensor product of an operator on a
1-bit or 2-bit subspace of the Hilbert space $\mathcal{H}$ and
identity operators for the other factors of the tensor product.  Note
that each permutation of a non-identity factor with an identity factor
is a distinct gate, calling for a distinct command to the instruments
that implement it.  For this reason, the number of quantum gates for
an $n$-bit quantum computer grows faster than $n$.  Call this number
$G(n)$ and let the set of gates be $U_1, \ldots, U_G$.  The most
commonly used models of quantum computers can be put in the form
\cite{deutsch2}:
\begin{itemize} \item Prepare a starting state independent of the
input ({\em e.g.}\ the integer to be factorized). \item Transform the
state by a product of quantum gates that depends on the input.
\item Make a measurement independent of the input. \end{itemize}

For an example, suppose the scientist assumes properties of models 1
through 4 and looks for the model that gives the least mean-square
deviation between relative frequencies of outcomes and probabilities
calculated by (\ref{eq:basic2}).  To factorize an integer $I$, a
classical computer program is converted to a product of $K(I)$ quantum
gates, a number that rises faster than linearly with $\log I$.  To
obtain the effect of multiplying the gate transformations, the
scientist must first have solved the model to determine the command
$b_{U,j}$ for each gate $U_j$ occurring in the product.  As in the
portrait in section 3 of putting tokens into a net fragment, the
scientist programs a CPC to transmit a command $b_v$ to prepare an
initial state $|v\rangle$, commands $b_{U,j}$ for the gates needed,
and a command $b_M$ for a measurement.  This endeavor is known to
exhibit the following four features:
\begin{enumerate}

\item The instruments are valuable as a quantum computer insofar as
their results substitute for a more costly classical calculation
defined by the model.

\item An inexpensive classical computation ({\em e.g.}\ with the CPC) tests
whether outcomes interpreted from results correctly solve the problem.

\item Quantum indeterminacy imposes a positive probability that a
result fails to provide a correct answer, so multiple tries with the
instruments are the rule, and a wrong answer does not by itself imply
a fault in the instruments.

\item The tolerable imprecision of instruments implementing the chosen
model of a quantum gate diminishes as the inverse of the number of gates
$K(I)$ in the sequence \cite{vaz}. \end{enumerate}

\begin{figure}[tb]
\begin{center}
\epsfbox{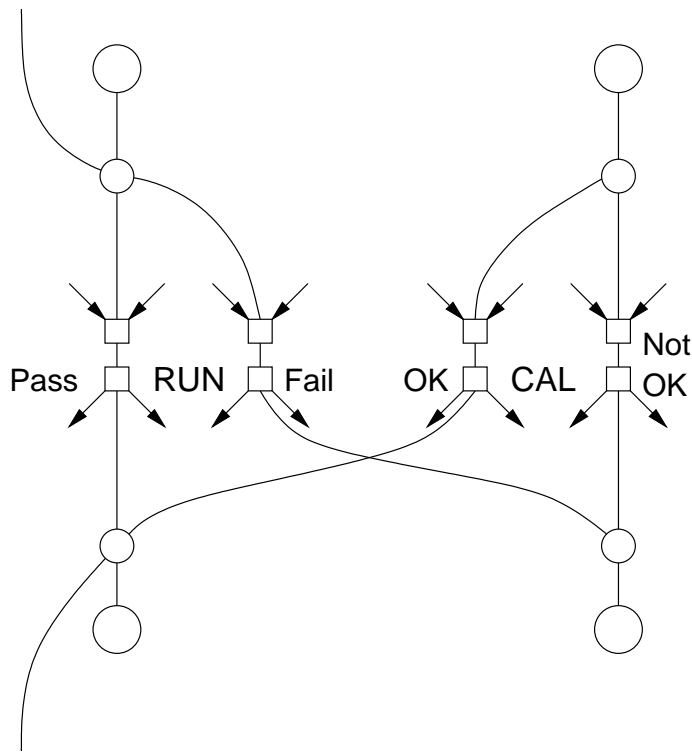} 
\end{center}
\caption{Alternating between running and testing a QC.}
\label{fig:modes}\end{figure}

Because the number of gates required in the product rises with the
size of the integer to be factorized, feature 4 implies that passing
the test for smaller integers is no guarantee against failure of the
instruments to factorize larger integers, unless the model or the
instruments or both are refined.  This requires, in turn, that a CPC
intended for use on progressively larger integers be organized to
switch between a mode of using the quantum computer and a mode of
inquiring into its performance, {\em e.g.}\ so as to determine commands
that make it behave more precisely in accord with the desired quantum
gates.  This calls for a program for the CPC that expands the events
``Use existing program'' of figure \ref{fig:utmp2} to that of figure
\ref{fig:modes}.

\subsection{Navigating the lattice of models to get better commands}

As an example of what goes on within the coarsely portrayed event
``Calibrate,'' suppose a scientist who uses a model $\alpha$ of the
form $(|v\rangle, U, M)_B$ finds it works for small integers, but
fails for bigger ones, which require more precise gates, which in turn
requires calibrating ({\em i.e.}\ adjusting) the commands used to
generate gates.  This means giving up model $\alpha$ and choosing some
alternative model $\beta$.  A scientist does not choose a model all at
once, but starts with some set of models and then narrows down on a
smaller set, sometimes to a single model, a process open to guesswork
at various stages.  At one stage, the scientist may need to relax a
constraint on models, leading to a bigger set of models from which to
choose; at another stage the scientist may guess a new constraint,
narrowing the set of models under consideration.  By such a back and
forth procedure, the scientist gives up $U_\alpha$ and arrives at a
new function $U_\beta$ (and hence a new model) with the hope that
solving this function for a command $b_{U,j,\beta}$ for gates $U_j$,
$j = 1, 2 \dots$, that will succeed for factorizing larger integers
than did the commands obtained from $U_\alpha$.  (This makes a need
for models adapted to homing in on results, with some metric on B, so
that a small change in the command $b_U$ results in a small change in
{\em e.g.}\ $U(b_U)$; while properties 3 and 4 are a start, going
beyond them is left to the future.)

To get a better model, the scientist guesses a set of models and hopes
to find within it a model that better fits measured results
interpreted as outcomes.  If no model of the set adequately fits
these outcomes, the scientist can first broaden the set of models and
next try to guess a property that will narrow the set, not to the
original model, but to one that fits better.  The recognition of
guesswork assures us that so long as progressively more ambitious
goals of precision keep being introduced, there is no end to the need
for adjusting both models and the laboratory instruments.

\subsection{Sample sizes needed to choose between models of\break gates}

As discussed in section 2.5, the number of trials needed to
statistically distinguish one model from another is bounded from below
by the inverse square of a weighted statistical distance between the
two models.  Small numbers of experimental results can sometimes
decide between distant models, but never between models that are close.
In particular, distinguishing experimentally between two models for
quantum gates can demand large samples:

\begin{prop} Models $\alpha$ and $\beta$ that differ only in $U$, with
spectral norm $\parallel U_\alpha(b_U) - U_\beta(b_U) \parallel =
\epsilon > 0$, are statistically indistinguishable for a command $b$
unless
\begin{equation} N(b) \geq \epsilon^{-2}. \end{equation}
\end{prop}
\begin{proof}
The models $\alpha$ and $\beta$ under the stated condition are
unitarily equivalent to a pair of models that differ only in
$|v\rangle$ with $\cos|\langle v_\alpha | v_\beta \rangle |$ $\leq
\epsilon$.  The proposition then follows from (\ref{eq:dist}) and
(\ref{eq:dangle}).\end{proof}

We argue elsewhere that this is a serious and heretofore unappreciated
challenge to bringing instruments into working order as quantum
computers, made visible by attention to the need for guesswork in
linking of laboratory instruments to equations of quantum
mechanics \cite{spie}.  

\section{Concluding remarks}

G\"odel proved that no one true structure could be generated by
sitting in a room with blinds drawn, writing down axioms.  Quantum
mechanics tells us that with the blinds up and the world of physical
measurement available, the situation remains much the same.  Just as
the openings for new axioms are uncloseable in mathematical logic, so
in physics guesswork is part of the foundation.

The net formalism can be put to use both to address improving the
contacts between equations and instruments, fostering advances in
theory and in instrumentation, and, at a more abstract level, to pose
problems pertaining to universal Turing machines adapted to process
control.  By formalizing commands to instruments, the techniques
presented here extend the reach of set-based mathematics into the area
of contact between equations and instruments, and open to study within
physics of some of what physicist do in the course of doing physics.
This extends a parallel beachhead established already in mathematics
by G\"odel's study of what a mathematician does to prove a theorem and
Turing's analysis of a mathematician who makes a note by which to
resume an interrupted computation.

\section{Acknowledgment}
We acknowledge Amr Fahmy for showing us our debt to G\"odel's proof of
incompleteness in mathematical logic.  We are indebted to Steffen
Glaser, Raimund Marx, and Wolfgang Bermel for introducing us to the
subtleties of laboratory work aimed at nuclear-magnetic-resonance
quantum computers \cite{marx}.  To David Mumford we owe our
introduction to quantum computing from the standpoint of pure
mathematics.  We are greatly indebted to Anatol W.  Holt and to
C. A. Petri for conversations years ago, in which each pointed in his
own way to the still mysterious expressive potential of nets.

\end{document}